\documentclass[twocolumn,amsmath,amssymb,prd]{revtex4} 

\usepackage{graphicx} 
\usepackage{epsfig,epsf,graphics,psfrag} 
\usepackage[utf8x]{inputenc} 
\usepackage{times} 
\usepackage{float} 
\usepackage{color} 
\usepackage{amsfonts,amssymb,stmaryrd,latexsym,amsmath} 
\usepackage{multirow} 
\usepackage{array} 
\usepackage{mathrsfs} 
\usepackage{siunitx} 
\usepackage{booktabs} 
\usepackage{enumerate}
\usepackage{subfigure} 
\usepackage{hhline} 
\usepackage{epstopdf} 
\usepackage{slashed} 
\usepackage{bm}
\usepackage{hyperref}
\usepackage{cleveref}
\usepackage{ulem}

\begin{document}

\title{New spectrum of charm-strange meson with constituent quark model $c\bar{s}$ contributions}

\author{Liu-Lin~Wang}
\author{Xing-Meng~Zhao}
\author{Xiao-Hai~Liu}\email{xiaohai.liu@tju.edu.cn}\affiliation{ 
Center for Joint Quantum Studies and Department of Physics, School of Science, Tianjin University, Tianjin 300350, China
}
\author{Mao-Jun~Yan}\email{yanmj0789@swu.edu.cn}
\affiliation{
School of Physical Science and Technology, Southwest University, Chongqing 400715, China 
}

\date{\today}

\begin{abstract}

We systematically investigate the $S$-wave interactions between Nambu-Goldstone bosons (NGBs) and charmed mesons in the $(S,I)=(1,0)$ sector using the chiral unitary approach. The scattering amplitudes incorporate both the Weinberg-Tomozawa term and additional contributions from $s$- and $u$-channel exchanges of $c\bar{s}$ states predicted by the constituent quark model (CQM). Through analytic continuation of the unitarized amplitudes to the complex energy plane, we identify multiple poles corresponding to bound states and resonances. Our analysis reveals a rich spectrum of $D_{sJ}$ states across $J^P = 0^+, 1^+, 1^-$, and $2^-$ sectors, providing new insights into the nature of established resonances like $D_{s0}^*(2317)$ and $D_{s1}(2460)$, while predicting several new states that could be observed in future experiments.
\end{abstract}

\maketitle

\section{Introduction}

Quantum chromodynamics (QCD) serves as the fundamental theory of strong interactions, exhibiting non-perturbative behavior in the low-energy regime where color confinement allows diverse configurations to contribute to hadron spectroscopy. Understanding hadron spectroscopy is therefore crucial for unveiling the features of QCD in the non-perturbative domain. Following the discovery of $J/\psi$ \cite{E598:1974sol, SLAC-SP-017:1974ind}, the Cornell potential model was introduced to describe the charmonium spectrum \cite{Eichten:1974af}. Subsequent refinements, including fine interactions, led to the Godfrey-Isgur (GI) model \cite{Godfrey:1985xj, Capstick:1986ter}, which has been widely successful in explaining the spectra of mesons ($q\bar{q}$) and baryons ($qqq$) with $q=u,\,d,\,s,\,c,$ and $b$ quarks.
However, predictions from the GI model \cite{Godfrey:1986wj} exhibit significant discrepancies for $D_{s0}^{\ast}(2317)$ \cite{BaBar:2003oey} and $D_{s1}(2460)$ \cite{CLEO:2003ggt}. Specifically, the mass of $D_{s0}^{\ast}(2317)$ ($D_{s1}(2460)$) is lower than (nearly equal to) its non-strange counterpart $D_0(2400)$ ($D_1(2430)$). Further details on these discrepancies can be found in reviews \cite{Guo:2017jvc,Olsen:2017bmm,Liu:2019zoy,Chen:2016qju} and references therein.

Notably, the mass gap $(M_{D_{s1}(2460)}-M_{D_{s0}^{\ast}(2317)})$ is nearly equal to $(M_{D^{\ast}}-M_D)$. This observation suggests that $D_{s0}^{\ast}(2317)$ and $D_{s1}(2460)$ can be dynamically generated as bound states of the $DK$ and $D^{\ast}K$ systems, respectively. In other words, these states may be interpreted as hadronic molecules composed of $DK$ and $D^{\ast}K$. The molecular picture naturally accounts for the lower masses of $D_{s0}^{\ast}(2317)$ and $D_{s1}(2460)$ and their unusual mass splitting. This interpretation is further supported by lattice QCD simulations \cite{Liu:2012zya,Mohler:2013rwa}.

Very recently, the Belle-II collaboration reported the radiative decay of $D^{\ast}_{s0}(2317)$~\cite{Belle-II:2025dzk}. The measured ratio $\mathcal{B}\left(D_{s 0}^*(2317)^{+} \rightarrow D_s^{*+} \gamma\right) / \mathcal{B}\left(D_{s 0}^*(2317)^{+} \rightarrow D_s^{+} \pi^0\right)$ shows a slight deviation from predictions based on a pure molecular scenario \cite{Fu:2021wde}. This implies that $D_{s 0}^*(2317)$ may be an admixture of a pure $c \bar{s}$ state and a molecular component \cite{Belle-II:2025dzk}. Such configuration mixing is naturally expected in QCD and has been extensively studied in the charm-strange meson spectrum \cite{Bicudo:2005de, MartinezTorres:2011pr, Mohler:2013rwa, MartinezTorres:2014kpc, Ortega:2016mms, Albaladejo:2018mhb,Yang:2021tvc, Zhang:2024usz, Shen:2025qpj, Zhou:2020moj}.

Previous studies have presented different interpretations regarding the $D_{s0}^{\ast}(2317)$ spectrum: some approaches predict a single pole, while others suggest the presence of two poles in the charm-strange spectrum with $J^P=0^+$, where the lower pole would correspond to the physical $D_{s0}^{\ast}(2317)$ and a higher pole might represent a state not yet clearly identified. This discrepancy motivates two key questions: (i) How many poles exist in the charm-strange meson spectrum with $J^P=0^+$ in the relatively lower energy region? (ii) If two poles indeed exist, could the higher pole leave identifiable signatures in the $DK$ invariant mass distribution?

To address these questions, we reanalyze the mixing between $c\bar{s}$ states and molecular components within the chiral unitary approach. The $c\bar{s}$ states are taken from quark model predictions and coupled to meson-meson channels, while the contact-range interaction of meson pairs is determined by leading-order chiral perturbation theory. We extend our study to the $J^P=1^+,\,1^-$, and $2^-$ sectors, where the charm-strange spectrum with negative parities is also explored, including $c\bar{s}$ states and yielding richer structures than those predicted by quark models \cite{Godfrey:1985xj,Zeng:1994vj,Lahde:1999ih,DiPierro:2001dwf,Ebert:2009ua, Li:2010vx,Godfrey:2015dva,Ni:2023lvx}. Moreover, our predictions in the $J^P=1^-$ sector with mass around 2.8 $\rm{GeV}$ provide new insights into the nature of $D_{s1}(2860)$ \cite{ParticleDataGroup:2024cfk}.

This paper is organized as follows. In Sec.~\ref{sec:Framework}, we construct the effective Lagrangian for Nambu–Goldstone bosons (NGBs) scattering off charmed mesons and unitarize the amplitude within the unitary chiral approach. Numerical results for dynamically generated poles are presented in Sec.~\ref{sec:result}. A brief summary is given in Sec.~\ref{sec:summary}.

\section{Theoretical Framework}
\label{sec:Framework}
\subsection{Chiral potentials}

The low-energy $S$-wave interactions between NGBs and charmed mesons are described within heavy meson chiral perturbation theory (HM$\chi$PT). In this effective field theory, the mass and field gradients of heavy mesons are counted as $\mathcal{O}(p^0)$, while those of NGBs are counted as $\mathcal{O}(p^1)$, where $p$ denotes the expansion parameter.

The leading-order kinetic Lagrangian for S-wave heavy-light mesons reads \cite{Wise:1992hn,Casalbuoni:1996pg}
\begin{equation}
    \label{kineticLag}
     \mathcal{L}_{LO}=i\langle\bar{H_a}v_{\mu}\mathcal{D}_{ba}^{\mu}H_b\rangle,
\end{equation}
with the heavy superfield definded as
\begin{equation}
    \label{heavyfield}   
    \begin{aligned}
        H_a=&\frac{1+\slashed{v}}{2}[P^{\ast\mu}_{a}\gamma_{\mu}-P_a\gamma_5],\\ \bar{H}_a=&\gamma_0H^{\dagger}_a\gamma_0,\\
    \end{aligned}
\end{equation}
where $v$ denotes the heavy meson velocity and $\langle \cdots \rangle$ represents the trace over Dirac matrices. The flavor index $a$ labels light quarks. The $H_a$ doublet contains $P = (D^0, D^+, D_s^+)$ and $P^{\ast} = (D^{\ast 0}, D^{\ast +}, D_{s}^{\ast +})$ with $s_\ell^p = \tfrac{1}{2}^-$,  where $s_\ell^p$ represents the total angular momentum and parity of the light degrees of freedom in the heavy quark limit. The covariant derivative is 
\begin{equation}
    \label{cd}
    \mathcal{D}_{\mu}P_{a}=\partial_{\mu}P_{a}-\Gamma_{\mu}^{ba}P_{b},\quad\mathcal{D}^{\mu}P_{a}^{\dagger}=\partial^{\mu}P_{a}^{\dagger}+\Gamma_{ab}^{\mu}P_{b}^{\dagger},
\end{equation}
with the chiral connection 
\begin{equation}
    \label{cc}
    \Gamma_\mu=(u^\dagger\partial_\mu u+u\partial_\mu u^\dagger)/2,
\end{equation}
where
\begin{equation}
    \label{NBGs}
    u=\mathrm{exp}\left(\frac{i\Phi}{\sqrt{2}F_0}\right),
\end{equation}
and $F_0 = 92.4\,\rm{MeV}$ is the pion decay constant. The NGB octet matrix is
\begin{equation}
    \Phi=\begin{pmatrix}
        \tfrac{1}{\sqrt{2}}\pi^0+\tfrac{1}{\sqrt{6}}\eta&\pi^+&K^+\\
        \pi^-&-\frac{1}{\sqrt{2}}\pi^0+\tfrac{1}{\sqrt{6}}\eta&K^0\\
        K^-&\bar{K}^0&-\tfrac{2}{\sqrt{6}}\eta
        \end{pmatrix}.
\end{equation}

Substituting Eqs.~(\ref{heavyfield})--(\ref{NBGs}) into Eq.~(\ref{kineticLag}) yields the Weinberg-Tomozawa (WT) term
\begin{equation}
	\mathcal{L}_{\text{WT}} = \frac{i}{4F_0^2} \langle \bar{H}_a v_\mu [\Phi, \partial^\mu \Phi]_{ba} H_b \rangle.
\end{equation}
	
For $D_i(p_1)\Phi_i(p_2) \to D_f(p_3)\Phi_f(p_4)$ scattering, the contact potential is
\begin{equation}
	\label{0+chiral potential}
	V_c^{0^+} = C_0 \frac{s-u}{4F_0^2},
\end{equation}
with Mandelstam variables $s=(p_1+p_2)^2$, $t=(p_1-p_3)^2$, $u=(p_1-p_4)^2$. For $(S,I)=(1,0)$ channels, the coefficient matrix is
\begin{equation}
	C_0 = 
    \begin{pmatrix}
		-2 & -\sqrt{3} \\
		-\sqrt{3} & 0
	\end{pmatrix},
\end{equation}
where channels 1 and 2 correspond to $DK$ and $D_s\eta$, respectively.
	
Heavy quark spin symmetry (HQSS) implies similar potentials for $D^*\phi$ scattering with an additional factor $\varepsilon(p_1)\cdot\varepsilon^*(p_3)$, with $\varepsilon$ being the polarization vector. In the heavy-quark limit, $\varepsilon(p_1)\cdot\varepsilon^*(p_3) = -1$ \cite{Abreu:2011ic}.
	
For $s_\ell^p = \tfrac{3}{2}^+$ $P$-wave heavy-light mesons, the WT term becomes
\begin{equation}
	\label{TkineticLag}
	\mathcal{L}_{\text{WT}} = \frac{i}{4F_0^2} \langle \bar{T}_{a\nu} v_\mu [\Phi, \partial^\mu \Phi]_{ba} T_b^\nu \rangle,
\end{equation}
with the superfield
\begin{eqnarray}
	\label{Theavyfield}
	T_a^\mu = &&\frac{1+\slashed{v}}{2} \Bigg\{
	P_{2a}^{\ast\mu\nu}\gamma_\nu
	- \sqrt{\frac{3}{2}} P_{1a\nu}
	\nonumber \\
    &&\times \gamma_5 \big[g^{\mu\nu} - \frac{1}{3}\gamma^\nu(\gamma^\mu - v^\mu)\big]
		\Bigg\},
\end{eqnarray}
and $\bar{T}_a^\mu = \gamma_0 T_a^{\mu\dagger} \gamma_0$, where $P_{1} = (D_1^0, D_1^+ ,D_{s1}^+)$ and $P_{2}^{\ast} =(D^{\ast0}_{2}, D^{\ast+}_{2} ,D_{s2}^{\ast+})$. Chiral symmetry ensures similar contact potential forms for $J^P = 1^-$ and $2^-$ channels \cite{Hyodo:2006kg}.

\subsection{Bare state exchange potentials}

\begin{figure*}[t]
  \centering
  \includegraphics[width=0.8\linewidth]{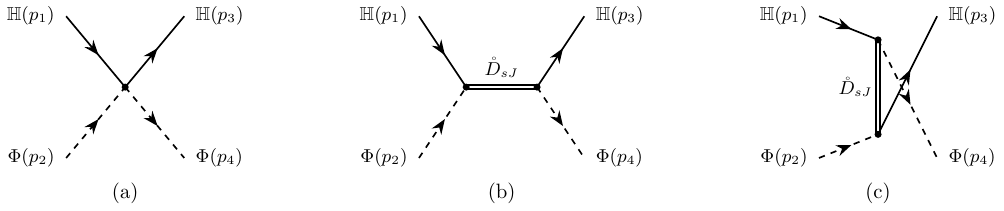}
  \caption{Distinct interaction channels for NGB--charmed meson scattering: (a) Weinberg-Tomozawa term, (b) $s$-channel pole from a bare $c\bar{s}$ state, and (c) $u$-channel exchange.}
  \label{fig:HK}
\end{figure*}

In the constituent quark model (CQM), several $c\bar{s}$ excitations are predicted with masses near the thresholds of NGBs and charmed mesons. The potential mediated by the exchange of these CQM $c\bar{s}$ states is expected to contribute significantly to the $\mathbb{H}\Phi$ coupled-channel interactions, as illustrated in Fig.~\ref{fig:HK}. Here, $\mathbb{H}$ collectively denotes the heavy charmed meson fields. We refer to the CQM states that lack coupled-channel interactions as bare states. 

In the $J^P = 0^+$ sector, the mass of the CQM $1^3P_0$ state lies above the $DK$ threshold. Higher radial excitations $n^{3}P_0$ ($n \ge 2$) are not considered in this work, as their masses fall well beyond the energy region of interest.

The situation becomes more involved in the $J^P = 1^+$ sector. Here, the relevant CQM states are $1P_1$ and $1P_1'$, which undergo configuration mixing described by 
\begin{equation}
\label{mix}
\begin{aligned}
    L^{\prime} &= {}^1L_J\cos\theta_L + {}^3L_J\sin\theta_L ,\\
    L &= -{}^1L_J\sin\theta_L + {}^3L_J\cos\theta_L ,
\end{aligned}
\end{equation}
where $L = P$ (i.e., $L = 1$) for $P$-wave mixing. An analogous mixing scheme applies to states with higher orbital angular momenta, such as $D$- and $F$-waves, under the relation $J = L$.

In the heavy quark limit, the broad $1P_1'$ state couples to $D^*K$ via $S$-wave, while the narrow $1P_1$ state couples via $D$-wave. For the $J^P = 1^-$ sector, the corresponding CQM states are $1^3D_1$ ($s_\ell^p=\tfrac{3}{2}^-$) and $2^3S_1$ ($s_\ell^p=\tfrac{1}{2}^-$). Under the HQSS, the $1^3D_1$ state exhibits strong $S$-wave coupling to $D_1 K$, while the $S$-wave coupling of $2^3S_1$ with $D_1 K$ is highly suppressed. 
A similar structure appears in the $J^P = 2^-$ sector. Here, the $1D_2'$ ($s_\ell^p = \tfrac{3}{2}^-$) couples to $D_2^*K$ in $S$-wave, while the $1D_2$ ($s_\ell^p = \tfrac{5}{2}^-$) couples in $D$-wave. In the heavy quark limit, we have the mixing angles $\theta_P = -54.7^\circ$ and $\theta_D = -50.8^\circ$ for $P$-wave and $D$-wave states, respectively~\cite{Godfrey:1986wj,Cahn:2003cw}.

The effective potential for the $s$-channel exchange of the bare state $\mathring{D}_{sJ}$ is given by  
\begin{eqnarray}
    \label{vs}
    V_{s}^{0^+}&=&C_s^{0^+}\frac{1}{s-\mathring{m}_{c\bar{s}}^2} ,
    \\
    V_{s}^{1^+}&=&C_s^{1^+}\varepsilon_{\mu}(p_1)\frac{-g^{\mu\nu}+v^{\mu}v^{\nu}}{s-\mathring{m}_{c\bar{s}}^2}\varepsilon_{\nu}^{\ast}(p_3) ,
    \\
    V_{s}^{1^-}&=&C_s^{1^-}\varepsilon_{\mu}(p_1)\frac{-g^{\mu\nu}+v^{\mu}v^{\nu}}{s-\mathring{m}_{c\bar{s}}^2}\varepsilon_{\nu}^{\ast}(p_3),
\\
    V_{s}^{2^-}&=&C_s^{2^-}\varepsilon_{\mu\nu}(p_1)\frac{\mathcal{P^{\mu\nu\alpha\beta}}}{s-\mathring{m}_{c\bar{s}}^2}\varepsilon_{\alpha\beta}^{\ast}(p_3),
\end{eqnarray}
where the spin-2 projection operator is defined as
\begin{eqnarray}
&&    \mathcal{P_{\mu\nu\alpha\beta}}=\frac{1}{2}(\tilde{g}_{\mu\alpha}\tilde{g}_{\nu\beta}+\tilde{g}_{\mu\beta}\tilde{g}_{\nu\alpha})-\frac{1}{3}\tilde{g}_{\mu\nu}\tilde{g}_{\alpha\beta},
\nonumber\\
&&    \tilde{g}_{\mu\nu}=-g_{\mu\nu}+v_{\mu}v_{\nu}.
\end{eqnarray}
Here, $\mathring{m}_{c\bar{s}}$ denotes the mass of the bare state $\mathring{D}_{sJ}$, and $C_s$ represents the coupling matrix between $\mathring{D}_{sJ}$ and the $\mathbb{H}\Phi$ system, which depends on the specific $J^P$ sector.
Additionally, the $u$-channel exchange potential is also taken into account in our model, whose form is obtained by substituting $s \to u$ and $C_s \to C_u$ in the above expressions.

\subsection{Unitarized amplitude}

To identify possible dynamically generated bound states and resonances, it is necessary to unitarize the amplitude obtained from the perturbation theory.  A commonly adopted approach is the Bethe--Salpeter equation, which in the on-shell approximation yields the unitarized scattering amplitude in coupled-channel matrix form as~\cite{Pelaez:2015qba,Oset:1997it}
\begin{equation}
    \mathcal{T}(s) = \left[1 - \mathcal{V}(s) G(s) \right]^{-1} \mathcal{V}(s),
\end{equation}
where $\mathcal{V}$ denotes the driving potential and $G$ is a diagonal matrix whose elements are the scalar one-loop two-meson propagators. Restricting to $S$-wave interactions, the potential is obtained by projecting the full amplitude as
\begin{equation}
    \mathcal{V}(s) = \frac{1}{2} \int_{-1}^{1} \left[ V_c(s,t,u) + V_s(s,t,u) + V_u(s,t,u) \right]  d\cos\theta,
\end{equation}
with $\theta$ being the scattering angle. The loop function for channel $j$ is given by
\begin{equation}
    G_j(s) = i \int \frac{d^4 q}{(2\pi)^4} \frac{1}{(q^2 - m_j^2 + i\epsilon)((P - q)^2 - M_j^2 + i\epsilon)},
\end{equation}
where $m_j$ and $M_j$ are the masses of the pseudoscalar meson and heavy meson, respectively. Using dimensional regularization, the loop function can be expressed as
\begin{equation}
    \begin{aligned}
        G^{\text{Dim.}}_j(s) &= \frac{1}{16\pi^2} \Bigg\{ a(\mu) + \ln \frac{M_j^2}{\mu^2} + \frac{s - M_j^2 + m_j^2}{2s} \ln\frac{m_j^2}{M_j^2} \\
        &+ \frac{\kappa_j(s)}{2s} \left[ \ln \left( \frac{s - m_j^2 + M_j^2 + \kappa_j(s)}{-s + m_j^2 - M_j^2 + \kappa_j(s)} \right) \right. \\
        &\left. + \ln \left( \frac{s + m_j^2 - M_j^2 + \kappa_j(s)}{-s - m_j^2 + M_j^2 + \kappa_j(s)} \right) \right] \Bigg\},
    \end{aligned}
\end{equation}
where $\mu$ is the renormalization scale and $\kappa_j(s) = \sqrt{ [s - (m_j + M_j)^2][s - (m_j - M_j)^2] }$. Alternatively, the loop function can be regularized via a three-momentum cutoff $q_{\text{max}}$, denoted as $G_j^{\text{Cut}}(s, q_{\text{max}})$, whose explicit form can be found in Ref.~\cite{Oller:1998hw}.

In the chiral unitary approach, dynamically generated states manifest as poles of the scattering T-matrix located on various Riemann sheets (RS). Bound states are identified with poles on the real axis below the lowest threshold on the first RS, while resonances are associated with complex poles on the second RS, located below the real axis. The pole positions of the bound state and the resonance are parametrized as 
\begin{equation}
    \sqrt{s^I_{\text{pole}}} = M_{\text{B}}, \quad \sqrt{s^{II}_{\text{pole}}} = M_R - i \frac{\Gamma_R}{2},
\end{equation}
where the real part corresponds to the mass of the state, and the imaginary part gives half the resonance width.

To quantify the molecular content of dynamically generated states, we employ the compositeness criterion~\cite{Weinberg:1962hj,Baru:2003qq}. For bound states, the probability of finding the molecular component in channel $j$ is given by~\cite{Gamermann:2009uq,Aceti:2014ala}
\begin{equation}
    P_j = -g_j^2 \left. \frac{\partial G_j}{\partial s} \right|_{s = M_{\text{B}}^2},
\end{equation}
where $j$ labels the hadronic channel, $M_{\text{B}}$ denotes the bound state mass, and $g_j$ represents the coupling to channel $j$. In the case of resonant states, the compositeness becomes complex-valued, and a rigorous probabilistic interpretation is not well-defined in such circumstances. Following established approaches~\cite{Aceti:2012dd}, we therefore adopt the modulus $|P_j|$ as an effective measure of the molecular component. Near a pole at $s_{\text{pole}}$, the scattering amplitude exhibits the characteristic behavior
\begin{equation}
    \mathcal{T}_{ij}(s) \simeq \frac{g_i g_j}{s - s_{\text{pole}}},
\end{equation}
where $i$ and $j$ denote channel indices. The coupling strength $|g_j|$ can thus be extracted from the residue of the diagonal amplitude $\mathcal{T}_{jj}$ at the pole position.

\section{Numeric result}
\label{sec:result}

\subsection{Model parameters}

This section summarizes the parameter values used in our calculations. The physical masses of external mesons are taken from the Particle Data Group (PDG)~\cite{ParticleDataGroup:2024cfk}. The bare $c\bar{s}$ state masses are adopted from the relativistic quark model calculation in Ref.~\cite{Godfrey:2015dva}, as listed in Table~\ref{tab:cqmDmass}.

\begin{table}[htbp]
\centering
\caption{Masses (in MeV) of $c\bar{s}$ mesons predicted by the quark model~\cite{Godfrey:2015dva}.}
\label{tab:cqmDmass}
\begin{tabular}{cccccccc}
\hline\hline
State & $1^3P_0$ & $1P_1$ & $1P_1^{\prime}$ & $1^3D_1$ & $2^3S_1$ & $1D_2$ & $1D_2^{\prime}$\\
\hline
Mass & 2484 & 2549 & 2556 & 2899 & 2732 & 2900 & 2926\\ 
\hline\hline
\end{tabular}
\end{table}

The coupling constants $g_{\mathring{D}_{sJ}\mathbb{H}\Phi}$ appearing in $C_s$ and $C_u$ are determined by matching the effective decay width to the chiral quark model prediction
\begin{equation}
    \Gamma_{\text{eff}}(g_{\mathring{D}_{sJ}\mathbb{H}\Phi}) = \Gamma_{\chi\text{QM}}.
\end{equation}
The detailed matching procedure is described in Appendix~\ref{cqm}, with the resulting numerical values provided in Table~\ref{tab:cqmcoupling}.

\begin{table}[htbp]
\centering
\caption{Effective coupling constants (in GeV) for bare state $\mathring{D}_{sJ}$--$\mathbb{H}\Phi$ vertices evaluated in the chiral quark model.}
\label{tab:cqmcoupling}
\begin{ruledtabular}
\begin{tabular}{lcccc}
State & Channel	& Coupling & Channel &Coupling\\\hline
$1^3P_0$ & $DK$ & $12$ & $D_s\eta$ & $6.1$\\
$1P_1$   & $D^{\ast}K$ & $0.87$ & $D^{\ast}_s\eta$ & $1.1$\\
$1P_1^{\prime}$ & $D^{\ast}K$ & $14$ & $D^{\ast}_s\eta$ & $6.7$\\
$1^3D_1$ & $D_1K$ & $22$ & $D_{s1}\eta$ & $11$\\
$2^3S_1$ & $D_1K$ & $3.9$ & $D_{s1}\eta$ & $3.6$\\
$1D_2$ & $D^{\ast}_2K$ & $0.69$ & $D^{\ast}_{s2}\eta$ & $1.4$\\
$1D_2^{\prime}$ & $D^{\ast}_2K$ & $22$ & $D^{\ast}_{s2}\eta$ & $11$\\
\end{tabular}
\end{ruledtabular}
\end{table}

We employ dimensional regularization for the loop functions, with the subtraction constant $a^{\mathbb{H}\Phi}(\mu=1~\mathrm{GeV})$ determined by matching to the cutoff-regularized scheme:
\begin{equation}
    \label{cutmatching}
    G_j^{\text{Dim.}}(s_{\text{th}}^{j}, a(\mu)) = G_j^{\text{Cut}}(s_{\text{th}}^{j}, q_{\text{max}}),
\end{equation}
where $s_{\text{th}}^j$ denotes the threshold of channel $j$. To reproduce the PDG value of the $D_{s0}^*(2317)$ mass, $2317.8~\text{MeV}$, we obtain 
$q_{\text{max}} = 775~\text{MeV}$ when using only the WT term, and 
$q_{\text{max}} = 550~\text{MeV}$ when including both the WT term and the bare-state exchange potential. This value of $q_{\text{max}}$ is subsequently employed consistently across all $\mathbb{H}\Phi$ scattering processes within our matching scheme, ensuring a unified treatment of the regularization procedure throughout our analysis.

\subsection{$D_{sJ}$ spectrum with $J^P=0^+$ and $1^+$}

\begin{table*}[t]
\caption{Pole positions, couplings to hadronic channels, and compositeness for dynamically generated states in the $J^P=0^+$ sector. The first column indicates the bare states included in the potential, with the corresponding subtraction constant $a(\mu)$ specified for each configuration.}
\begin{ruledtabular}
\begin{tabular}{c c c c c c c c }
  Bare state & $a$ & Pole position [MeV] & $g_{DK}$[GeV]  & $g_{D_s\eta}$[GeV] & $P_{DK}$[\%] & $P_{D_s\eta}$[\%]\\\hline
 -- & $-1.847$ & $2317.8$ & $10$ & $6.8$ & $63$ & $11$ \\\hline
\multirow{2}{*}{$1^3P_0$} & \multirow{2}{*}{$-1.553$} & $2317.8$ & $11$ & $6.5$ & $70$ & $10$ \\
 &  & $2579.6-15.1i$ & $4.1$ & $4.2$ & $5.2$ & $9.5$ \\
\end{tabular}
\end{ruledtabular}
\label{tab:charm0+}
\end{table*}

\begin{table*}[t]
\caption{Pole positions, couplings, and compositeness for dynamically generated states in the $J^P=0^+$ sector obtained with a Gaussian regulator ($\Lambda=800$ MeV).}
\begin{ruledtabular}
\begin{tabular}{ c c c c c c c}
 Bare state & $a$ & Pole position [MeV] & $g_{DK}$[GeV] & $g_{D_s\eta}$[GeV] & $P_{DK}$[\%] & $P_{D_s\eta}$[\%] \\\hline
 --       & $-1.774$ & $2317.8$           & $11$ & $9.1$ & $79$ & $20$ \\\hline
  \multirow{2}{*}{$1^3P_0$} & \multirow{2}{*}{$-1.549$} & $2317.8$           & $11$ & $7.9$ & $53$ & $17$ \\
   &  & $2654.2-84.8i$    & $9.3$ & $8.6$ & $21$ & $23$ 
\end{tabular}
\end{ruledtabular}
\label{tab:charm0+ff}
\end{table*}

In the $J^P = 0^+$ $D_{(s)}\Phi$ scattering, the WT interaction dynamically generates the $D_{s0}^*(2317)$ as a pole in the $DK$–$D_s\eta$ coupled-channel system, a result well established in previous studies~\cite{Hofmann:2003je,Guo:2006fu,Wang:2006zw,Faessler:2007gv,Guo:2008gp,Guo:2009ct,Liu:2012zya,Cleven:2014oka,Guo:2015dha,Albaladejo:2016hae,Du:2017ttu,Albaladejo:2018mhb,Yang:2021tvc,Fu:2021wde,Gil-Dominguez:2023puj}. As shown in the first row of Table~\ref{tab:charm0+}, the subtraction constant $a(\mu)$ serves as the sole free parameter, encoding short-range dynamics. With $a(\mu) = -1.847$, the extracted couplings and compositeness confirm that $D_{s0}^*(2317)$ is a dynamically generated state dominated by the $DK$ molecular component.

A bare $c\bar{s}$ state with mass $2484~\mathrm{MeV}$, predicted by the quark model~\cite{Godfrey:2015dva}, is included in the second row of Table~\ref{tab:charm0+}. This state lies between the $DK$ and $D_s\eta$ thresholds and couples to both $s$- and $u$-channel meson–meson scattering, interfering with the WT interaction. To reproduce the physical mass of $D_{s0}^*(2317)$, the subtraction constant shifts from $-1.847$ to $-1.553$, reflecting a moderate change in the effective potential near the $DK$ threshold. The resulting variations in couplings and compositeness remain perturbative, and the $DK$ molecular nature of $D_{s0}^*(2317)$ is preserved, in agreement with earlier works.

In addition to $D_{s0}^*(2317)$, a second pole emerges at $2579.6 - i15.1~\mathrm{MeV}$, with a width of about $30.2~\mathrm{MeV}$. This is consistent with the prediction in Ref.~\cite{Albaladejo:2018mhb}, though not noted in Ref.~\cite{Yang:2021tvc}. The pole is shifted by about $96~\mathrm{MeV}$ from the bare mass, a much larger effect than in the $\Xi_{cc}\bar{K}$–$\Omega_{cc}\eta$ system with a subthreshold bare pole, as studied in our earlier work~\cite{Yan:2018zdt}. This difference arises from the stronger coupling between the hadronic channels and the bare $c\bar{s}$ state in the $DK$–$D_s\eta$ case, where HQSS permits a sizeable $S$-wave coupling. In contrast, the $\Xi_{cc}\bar{K}$–$\Omega_{cc}\eta$ system involves a HQSS-breaking coupling that is numerically small.

Although the bare $c\bar{s}$ state $D_s(1^3P_0)$ has a large decay width into $DK$ of more than two hundred MeV~\cite{Godfrey:2015dva}, the higher pole obtained here turns out to be significantly narrow. This narrowing results from the interference between the WT term and the $s$-channel bare-state exchange potential $V_s$. As noted in lattice QCD analyses of $DK$ scattering~\cite{Albaladejo:2018mhb}, the higher pole occupies a larger phase space for $DK$ decay than the bare state, yet its effective coupling is reduced. This can be understood from the behavior of $V_s$: when $s > \mathring{m}_{c\bar{s}}^2$, the sign of $V_s$ flips, turning the interaction from attractive to repulsive. The repulsive $V_s$ interferes destructively with the attractive WT term, yielding an effective $s$-channel potential with a shifted mass and a suppressed effective coupling. The real part of the higher pole corresponds to this effective mass, and the small effective coupling leads to the observed narrow width.

\begin{figure}
  \includegraphics[width=0.8\linewidth]{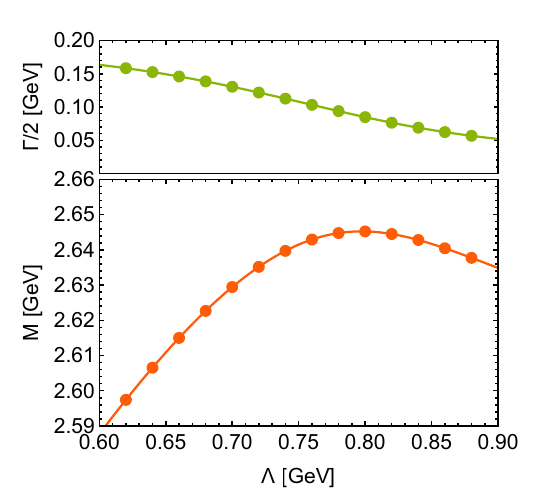}
\caption{Upper panel: width of the higher pole as a function of the regulator $\Lambda$. Lower panel: corresponding pole mass variation. The subtraction constant is fixed at $a(\mu) = -1.549$.}
  \label{fig:pole-position}
\end{figure}

The higher pole decays into $DK$ with an outgoing kaon momentum of approximately $440~\mathrm{MeV}$, which is comparable to our chosen cutoff $q_{\text{max}} = 550~\mathrm{MeV}$. As the $DK$ scattering energy increases, the validity of the WT term becomes questionable. To address this high-energy ambiguity, we introduce a Gaussian regulator to the WT term in the Bethe-Salpeter equation
\begin{equation}
    \tilde{V}_c=V_c \exp\left(-\frac{\bm{p}_2^2} {\Lambda^2}\right) \exp\left(-\frac{\bm{p}_4^2} {\Lambda^2}\right),
\end{equation}
where $\bm{p}_2$ and $\bm{p}_4$ represent the three-momenta of the initial and final light mesons in the center-of-mass frame, respectively.
The Gaussian regulator is introduced to address the issue of unnaturally narrow widths for the higher pole located above the $DK$ threshold, which is expected to exhibit stronger coupling to the $DK$ channel.
However, it is noted that a smaller $\Lambda$ would lead to a compositeness summation for the lower pole exceeding unity, thereby violating unitarity. Therefore, in the following discussion, we adopt a moderate value of $\Lambda = 800~\mathrm{MeV}$ for the Gaussian regulator to ensure unitarity preservation. The $\Lambda$-dependence of the higher pole position is shown in Fig.~\ref{fig:pole-position}. As $\Lambda$ varies from 600 to 900 MeV, the real part of the pole varies around 2.6 GeV, while the imaginary part decreases from 163 MeV to 52 MeV. With $\Lambda = 800~\mathrm{MeV}$ and $a(\mu) = -1.549$, the higher pole shifts to $(2654.2 - i84.8)~\mathrm{MeV}$ (see Table~\ref{tab:charm0+ff}). This represents a mass increase of approximately $75~\mathrm{MeV}$ compared to the unregulated case, while the width expands by nearly a factor of six. The enhanced width reflects a stronger effective coupling to the $DK$ channel when the Gaussian regulator is included.

It is worth noting that to reproduce the physical mass of $D_{s0}^*(2317)$ after introducing the Gaussian regulator, the cutoff $q_{\text{max}}$ in Eq.~(\ref{cutmatching}) is set to $702~\mathrm{MeV}$ for the WT interaction alone and to $547~\mathrm{MeV}$ when both WT and bare state interactions are included. The corresponding subtraction constants are adjusted to $a(\mu) = -1.774$ and $-1.549$, respectively. For the other states ($J^P = 1^+$, $1^-$, and $2^-$), we employ the same cutoff values within the Gaussian regulator framework as those established in the $0^+$ sector.

\begin{table*}[t]
\caption{Pole positions, couplings, and compositeness for the $J^P=1^+$ sector.} 
\begin{ruledtabular}
\begin{tabular}{ c c c c c c c }
 Bare state & $a$ & Pole position [MeV] & $g_{D^{\ast}K}$[GeV] & $g_{D_s^{\ast}\eta}$[GeV] & $P_{D^{\ast}K}$[\%] & $P_{D_s^{\ast}\eta}$[\%] \\\hline
 -- & $-1.932$ & $2459.3$ & $11$ & $7.2$ & $63$ & $11$ \\\hline
 \multirow{2}{*}{$1P^{\prime}_1$} & \multirow{2}{*}{$-1.654$} 
 & $2443.5$ & $12$ & $6.9$ & $64$ & $9.7$ \\
 & & $2672.6-22.1i$ & $5.4$ & $5.4$ & $9.4$ & $22$ \\\hline
 \multirow{3}{*}{$1P^{\prime}_1+1P_1$} & \multirow{3}{*}{$-1.654$} 
 & $2443.4$ & $12$ & $6.9$ & $64$ & $9.7$ \\
 & & $2673.1-22.2i$ & $5.5$ & $5.4$ & $9.4$ & $22$ \\
 & & $2549.0-0.0018i$ & $0.94$ & $2.0$ & $0.59$ & $1.2$ \\
\end{tabular}
\end{ruledtabular}
\label{tab:charm1+}
\end{table*}

\begin{table*}[t]
\caption{Pole positions, couplings, and compositeness for the $J^P=1^+$ sector with a Gaussian regulator ($\Lambda = 800~\mathrm{MeV}$). } 
\begin{ruledtabular}
\begin{tabular}{ c c c c c c c }
 Bare state & $a$ & Pole position [MeV] & $g_{D^{\ast}K}$[GeV] &  $g_{D_s^{\ast}\eta}$[GeV] & $P_{D^{\ast}K}$[\%] & $P_{D_s^{\ast}\eta}$[\%] \\\hline
 -- & $-1.863$ & $2460.0$ & $12$ & $9.7$ & $79$ & $20$ \\\hline
 \multirow{2}{*}{$1P^{\prime}_1$} & \multirow{2}{*}{$-1.650$} 
 & $2444.7$ & $12$ & $8.2$ & $64$ & $14$ \\
 & & $2727.4-77.7i$ & $11$ & $10 $ & $29$ & $37$ \\\hline
 \multirow{3}{*}{$1P^{\prime}_1+1P_1$} & \multirow{3}{*}{$-1.650$} 
 & $2444.6$ & $12$ & $8.2$ & $64$ & $14$ \\
 & & $2727.9-78.6i$ & $11$   & $10$ & $29$ & $37$ \\
 & & $2549.0-0.0017i$ & $1.2$ & $2.3$ & $0.93$ & $1.65$ \\
\end{tabular}
\end{ruledtabular}
\label{tab:charm1+ff}
\end{table*}

In the energy region of the higher pole, additional channels become relevant: $D_s\pi\pi$ and $D_s^*\pi\pi$ are open, while $D_s^* f_0(500)$ and $D^* \kappa(800)$ couple to the $J^P = 0^+$ sector in $P$-wave. Such $S$-$P$ wave coupled-channel effects can significantly enrich the spectrum \cite{Peng:2020gwk, Wu:2024bvl} but also complicate pole identification. Furthermore, the broad nature of $f_0(500)$ and $\kappa(800)$ suggests that three-body dynamics may couple non-negligibly to the $D_{(s)}\Phi$ system, though these effects are not included in our current framework. Consequently, the higher pole predicted in our coupled-channel $D_{(s)}\Phi$ scattering analysis may be difficult to observe in the $DK$ invariant mass distribution \cite{LHCb:2014ioa, BaBar:2014jjr, LHCb:2018oeb}.

HQSS establishes the $D^{\ast}_{(s)}$ mesons with $J^P=1^-$ as partners of the $D_{(s)}$ mesons, implying that the WT terms for NGBs scattering off $D^{\ast}_{(s)}$ are analogous to those in $D_{(s)}\Phi$ scattering. The short-range interactions, regulated by the subtraction constant $a(\mu)$ and parameterized by the cutoff $q_{\text{max}}$, are taken to be identical in both $D_{(s)}\Phi$ and $D^{\ast}_{(s)}\Phi$ systems. The numerical results are shown in Table~\ref{tab:charm1+}.
In this framework, the $D_{s1}(2460)$ is dynamically generated as a pole at $2459.3~\mathrm{MeV}$, consistent with previous studies~\cite{Guo:2006rp, Wang:2006zw,Gamermann:2007fi,Faessler:2007us,Cleven:2010aw,Cleven:2014oka,Du:2017zvv}.

In addition to this WT-driven pole, the quark model predicts two $c\bar{s}$ states---$1P_1$ and $1P_1'$---with masses of $2549~\mathrm{MeV}$ and $2556~\mathrm{MeV}$, respectively (see Table~\ref{tab:cqmDmass}). These correspond to mixed states of $^1P_1$ and $^3P_1$ configurations. Located between the $D^*K$ and $D_s^*\eta$ thresholds, they can couple to $D^*K$--$D_s^*\eta$ via $S$-wave, enriching the spectrum of charm-strange mesons in the $J^P=1^+$ sector.

When the WT interaction is included together with the $1P_1'$ bare state and $a(\mu)$ is set to $-1.654$, two poles emerge in the scattering amplitude. As shown in Table~\ref{tab:charm1+}, the lower pole is located at $2443.5~\mathrm{MeV}$, while the higher one appears at $2672.6 - i22.1~\mathrm{MeV}$. The proximity of the higher pole to the $D_s^*\eta$ threshold suggests significant interference between the bare state and the WT term. The narrow width of this pole results from a suppressed effective coupling to $D^*K$. The $S$-wave coupling between the $1P_1$ state and $D^*_{(s)}\Phi$ channels breaks HQSS, allowing its contribution to be treated perturbatively. When both $1P_1$ and $1P_1^\prime$ bare states are included simultaneously, three poles appear in the spectrum. Two of these poles exhibit masses slightly different from those in the single bare state ($1P_1'$) scenario, while the third pole remains close to the original bare state mass. This pattern indicates that the HQSS-breaking contributions play a relatively minor role in this configuration.

One may observe that the width of the higher pole at $2672.6~\mathrm{MeV}$ is relatively narrow, exhibiting behavior analogous to that found in the $0^+$ sector. Following the same procedure adopted for the $0^+$ sector, we introduce a Gaussian regulator to the WT term, which results in a substantial broadening of the width, as shown in Table~\ref{tab:charm1+ff}.

Based on the compositeness values, the lowest pole exhibits characteristics of a hadronic molecule in both cases (with and without the Gaussian regulator), while the other two states possess larger $c\bar{s}$ components.

It is also important to note that in the energy region around the higher pole, open channels such as $D_s \omega$ and $D K^*$ can couple in $S$-wave, while $D_s f_0(980)$ and $D \kappa(800)$ may contribute via $P$-wave. These channels, not included in the present model, could modify the properties of the higher pole, which may explain why it has not yet been clearly observed in experiment.

\begin{table*}[t]
\caption{Pole positions, couplings, and compositeness for dynamically generated states in the $J^P=1^-$ sector.} 
\begin{ruledtabular}
\begin{tabular}{ c c c c c c c }
 Bare state & $a$ & Pole position [MeV] & $g_{D_1K}$[GeV] & $g_{D_{s1}\eta}$[GeV] & $P_{D_1K}$[\%] & $P_{D_{s1}\eta}$[\%] \\\hline
 -- & $-2.171$ & $2874.5$ & $12$ & $8.5$ & $65$ & $12$ \\\hline
 \multirow{2}{*}{$1^3D_1$} & \multirow{2}{*}{$-1.930$} & $2826.3$ & $15$ & $8.1$ & $63$ & $9.7$ \\
  &  & $3092.6-31.1i$ & $7.2$ & $7.3$ & $12$ & $37$ \\\hline
 \multirow{3}{*}{$1^3D_1+2^3S_1$} & \multirow{3}{*}{$-1.930$} & $2826.2$ & $15$ & $7.9$ & $63$ & $9.2$ \\
  &  & $2735.5$ & $1.8$ & $2.6$ & $0.60$ & $0.96$ \\
  &  & $3094.4-31.7i$ & $7.2$ & $7.3$ & $12$ & $27$ \\
\end{tabular}
\end{ruledtabular}
\label{tab:charm1-}
\end{table*}

\begin{table*}[t]
\caption{Pole positions, couplings, and compositeness for dynamically generated states in the $J^P=1^-$ sector with a Gaussian regulator ($\Lambda = 800~\mathrm{MeV}$).} 
\begin{ruledtabular}
\begin{tabular}{ c c c c c c c }
 Bare state & $a$ & Pole position [MeV] & $g_{D_1K}$[GeV] & $g_{D_{s1}\eta}$[GeV] & $P_{D_1K}$[\%] & $P_{D_{s1}\eta}$[\%] \\\hline
 -- & $-2.111$  & $2877.6$ & $13$ & $11$ & $79$ & $22$ \\\hline 
 \multirow{2}{*}{$1^3D_1$} & \multirow{2}{*}{$-1.926$} & $2829.5$ & $14$ & $9.5$ & $62$ & $13$ \\
  &  & $3160.6-136.5i$ & $13$ & $12$ & $30$ & $33$ \\\hline
 \multirow{3}{*}{$1^3D_1+2^3S_1$} & \multirow{3}{*}{$-1.926$} & $2830.3$ & $14$ & $9.3$ & $61$ & $13$ \\
  &  & $2735.5$ & $1.3$ & $2.6$ & $0.31$ & $0.97$ \\
  &  & $3160.9-141.2i$ & $13$ & $12$ & $29$ & $32$ \\
\end{tabular}
\end{ruledtabular}
\label{tab:charm1-ff}
\end{table*}

\subsection{$D_{sJ}$ spectrum with $J^P=1^-$ and $2^-$}

For the NGBs scattering off the charm meson with $s_l^p = \frac{3}{2}^+$, the dynamics is similar to that in the $D^{(*)}_{(s)}\Phi$ scattering and contributes to the spectrum with $J^P = 1^-$ and $2^-$. Therefore, we extend the study to the $D_{sJ}$ states with $J^P = 1^-$ and $2^-$.

First, when only the WT terms are considered in the $D_{(s)1}\Phi$ scattering, a bound state is produced with a mass of $2874.5\,\mathrm{MeV}$, where the subtraction constant in the loop function is $a(\mu) = -2.171$, matched to $q_{\mathrm{max}} = 550\,\mathrm{MeV}$ in Eq.~(\ref{cutmatching}). 

Second, we examine the interference between the WT terms and the bare $c\bar{s}$ states. In the $J^P = 1^-$ sector, the quark model predicts two excited $c\bar{s}$ states that couple to $D_{(s)1} \Phi$ scattering via S-wave. As indicated in Table~\ref{tab:cqmDmass}, the mass of the $1^3D_1$ state is very close to that of the dynamically generated state obtained with only the WT terms ($2874.5$ MeV), suggesting that including this bare state in the potential could significantly alter the resulting spectrum. The results are shown in Table~\ref{tab:charm1-}.
When the $1^3D_1$ state is included, we use a subtraction constant $a(\mu) = -1.93$, and two poles emerge in the $D_{(s)1}\Phi$ scattering amplitude. The lower pole, located at $2826.3\,\mathrm{MeV}$, couples strongly to the $D_1 K$ channel and exhibits a sizable coupling to $D_{s1}\eta$. The presence of the bare $1^3D_1$ state pushes the WT-generated pole to a more deeply bound configuration, with a binding energy of approximately $90\,\mathrm{MeV}$. The higher pole is found at $3092.6\,\mathrm{MeV}$ with a width of $62.2\,\mathrm{MeV}$. Its mass is shifted by about 194 MeV from the bare mass of $2899\,\mathrm{MeV}$. This mass shift is notably larger than those observed in the $J^P = 0^+$ and $1^+$ sectors.

Finally, when both bare $c\bar{s}$ states are included together with the WT terms, three poles emerge in the $T$-matrix. The $2^3S_1$ state couples to $D_{(s)1}\Phi$ scattering in $S$-wave through HQSS-breaking interactions, and its interference with both the $1^3D_1$ state and the WT terms can be treated perturbatively. This is evidenced by the minimal mass shift of the lowest pole at $2735.5\,\mathrm{MeV}$ from the bare $2^3S_1$ mass of $2732\,\mathrm{MeV}$. Due to the HQSS-breaking nature of this coupling, the other two pole positions also show only slight modifications compared to the case with the $1^3D_1$ bare state alone.

As presented in Table~\ref{tab:charm1-ff}, the inclusion of a Gaussian regulator for the WT term leads to a broadening of the highest pole around 3.1 GeV, consistent with expectations. This broadening mainly arises from the enhanced coupling to the $D_1 K$ channel, which becomes comparable in strength to the $D_s(1^3D_1)$-$D_1K$ coupling listed in Table~\ref{tab:cqmcoupling}. 

Based on the compositeness value, the lowest pole can be interpreted as a $D_1 K$ bound state. Although our model does not explicitly calculate its decay width, one can expect that this $D_1 K$ molecular state can couple to channels such as $D^{(*)}K$, $D_s^* \pi\pi$, and $D^{(*)}K\pi$, suggesting that it would not be narrow and may correspond to the physical state $D_{s1}(2860)$.

The $D_{sJ}(2860)$ state was initially observed by the BaBar and LHCb collaborations in the $DK$ and $D^*K$ invariant mass spectra~\cite{BaBar:2006gme,BaBar:2009rro,LHCb:2012uts}. In 2014, LHCb performed a more detailed study of this state via the $B_s^0\to \bar{D}^0 K^- \pi^+$ decay process~\cite{LHCb:2014ott}. Their amplitude analysis indicates that the originally reported $D_{sJ}(2860)$ likely comprises two separate states: $D_{s1}(2860)$ with $J^P=1^-$ and $D_{s3}(2860)$ with $J^P=3^-$, with the following resonance parameters:
\begin{eqnarray}
M(D_{s1}(2860))&=&2859\pm 12 \pm 6 \pm 23 \ \mathrm{MeV}, \nonumber \\
\Gamma(D_{s1}(2860))&=&159\pm 23 \pm 27 \pm 72 \ \mathrm{MeV}, \nonumber \\
M(D_{s3}(2860))&=&2860.5\pm 2.6 \pm 2.5 \pm 6.0 \ \mathrm{MeV}, \nonumber \\
\Gamma(D_{s3}(2860))&=&53\pm 7 \pm 4 \pm 6 \ \mathrm{MeV} .
\end{eqnarray}
Considering these parameters, the $D_{s1}(2860)$ could be reasonably identified with the lowest pole obtained in our calculation, supporting its interpretation as a $D_1 K$ molecular state. For comprehensive reviews on these charm-strange mesons, we refer to Refs.~\cite{Guo:2017jvc, Chen:2016spr}.

\begin{table*}[t]
\caption{Pole positions, couplings, and compositeness for dynamically generated states in the $J^P=2^-$ sector.}
\begin{ruledtabular}
\begin{tabular}{ c c c c c c c }
 Bare state & $a$ & Pole position [MeV] & $g_{D_2^{\ast}K}$[GeV] & $g_{D_{s2}^{\ast}\eta}$[GeV] & $P_{D_2^{\ast}K}$[\%] & $P_{D_{s2}^{\ast}\eta}$[\%] \\\hline
 -- & $-2.193$ & $2913.0$ & $12$ & $8.6$ & $65$ & $12$ \\\hline
 \multirow{2}{*}{$1D_2^{\prime}$} &  \multirow{2}{*}{$-1.955$} & $2853.2$ & $15$ & $8.1$ & $59$ & $9.3$ \\
 &  & $3108.0-35.0i$ & $7.7$ & $7.6$ & $15$ & $26$ \\\hline
 \multirow{3}{*}{$1D_2^{\prime}+1D_2$} & \multirow{3}{*}{$-1.955$} & $2861.6$ & $15$ & $8.2$ & $62$ & $9.6$ \\
  &  & $2900.2$ & $0.11$ & $1.2$ & $0.042$ & $0.22$ \\
  &  & $3125.5-33.3i$ & $6.5$ & $7.0$ & $13$ & $17$ \\
\end{tabular}
\end{ruledtabular}
\label{tab:charm2-}
\end{table*}

\begin{table*}[t]
\caption{Pole positions, couplings, and compositeness for dynamically generated states in the $J^P=2^-$ sector with a Gaussian regulator ($\Lambda = 800~\mathrm{MeV}$).} 
\begin{ruledtabular}
\begin{tabular}{ c c c c c c c }
 Bare state & $a$ & Pole position [MeV] & $g_{D_2^{\ast}K}$[GeV] & $g_{D_{s2}^{\ast}\eta}$[GeV] & $P_{D_2^{\ast}K}$[\%] & $P_{D_{s2}^{\ast}\eta}$[\%] \\\hline
 -- & $-2.134$ & $2915.7$ & $13$ & $12$ & $79$ & $22$ \\\hline
 \multirow{2}{*}{$1D_2^{\prime}$} &  \multirow{2}{*}{$-1.951$} & $2853.7$ & $14$ & $9.3$ & $57$ & $12$ \\
 &  & $3165.8-123.6i$ & $14$ & $13$ & $37$ & $40$ \\\hline
 \multirow{3}{*}{$1D_2^{\prime}+1D_2$} & \multirow{3}{*}{$-1.951$} & $2865.6$ & $14$ & $9.4$ & $59$ & $13$ \\
  &  & $2900.2$ & $0.17$ & $1.23$ & $0.11$ & $0.24$ \\
  &  & $3190.7-134.1i$ & $13$ & $13$ & $31$ & $23$ \\
\end{tabular}
\end{ruledtabular}
\label{tab:charm2-ff}
\end{table*}

Given that the $D_2$ is the HQSS partner of the $D_1$, the study in the $J^P=1^-$ sector is naturally extended to the $J^P=2^-$ sector. 
In the $D_{(s)2}\Phi$ coupled-channel scattering, two bare $c\bar{s}$ states are located near the $D_2K$ threshold, with their masses listed in Tab.~\ref{tab:cqmDmass}. 
In the $S$-wave interaction, the $1D_2^{\prime}$ and $1D_2$ states couple to $D_2 K$ with HQSS conserved and broken, respectively.
Furthermore, the GI quark model predicts that the masses of $1^3D_1$ and $2^3S_1$ lie below the $D_1 K$ threshold, while those of $1D_2^{\prime}$ and $1D_2$ are below the $D_2 K$ threshold. 
Consequently, the behavior of the pole shifts obtained in $D_{(s)2}\Phi$ scattering exhibits similar patterns as those in $D_{(s)1}\Phi$ scattering.

The numerical results for the $J^P=2^-$ sector are summarized in Table~\ref{tab:charm2-}. When only the WT interactions are considered, a $D_2 K$ bound state is obtained at 2913 MeV. The inclusion of the bare $1D_2^{\prime}$ state shifts this WT-generated pole to a more deeply bound configuration.
When both the WT terms and the two bare states are included in the potential, three poles emerge in the scattering amplitude. The lowest pole, located at $2861.6\,\mathrm{MeV}$, couples strongly to the $D_2 K$ channel and can be interpreted as a $D_2 K$ molecular state. The other two poles, at $2900.2\,\mathrm{MeV}$ and $(3125.5-33.3i)\,\mathrm{MeV}$, are predominantly of $c\bar{s}$ character. Their masses are shifted from the bare $1D_2$ and $1D_2^{\prime}$ states by approximately $0.2\,\mathrm{MeV}$ and $200\,\mathrm{MeV}$, respectively. The minimal mass shift of the $1D_2$ state can be attributed to its HQSS-breaking coupling to the hadronic channels.

When a Gaussian regulator is introduced to the WT term, the pole positions undergo notable modifications: while the real parts exhibit only minor adjustments, the imaginary parts are substantially enhanced. These results, as detailed in Table~\ref{tab:charm2-ff}, indicate a significant increase in the widths of the poles, consistent with the strengthened couplings to the relevant hadronic channels.

\section{Summary}
\label{sec:summary}
In this work, we have systematically investigated the $S$-wave interactions between NGBs and charmed mesons in the $(S,I)=(1,0)$ sector within the chiral unitary approach. Our theoretical framework incorporates both the leading-order WT contact interactions and contributions from $s$- and $u$-channel exchanges of $c\bar{s}$ states predicted by the CQM, providing a comprehensive description of the charm-strange meson spectrum across multiple $J^P$ sectors.

In the $J^P = 0^+$ sector, the unitarized amplitude reveals two distinct poles. The lower pole is unambiguously identified with the established $D_{s0}^*(2317)$ resonance, which exhibits dominant $DK$ molecular character. The higher pole, located in the 2580--2650 MeV region, displays remarkable sensitivity to the regularization scheme. Particularly noteworthy is the effect of introducing a Gaussian regulator to the WT term: the higher pole undergoes significant broadening, with its width increasing from approximately 30 MeV in the unregulated case to more than 100 MeV, indicating enhanced effective couplings to the $DK$ channel.

HQSS naturally extends these considerations to the $J^P = 1^+$ sector, where the lower pole corresponds to $D_{s1}(2460)$, while a higher pole emerges around 2670--2730 MeV. The properties of this higher pole might be influenced by couplings to channels not included in our analysis, such as $D_s\omega$ and $DK^*$.

In the $J^P = 1^-$ sector, the interplay between HQSS-conserving bare states and WT terms suggests the presence of a lower pole around 2820--2870 MeV and a higher pole near 3090--3160 MeV. The $J^P = 2^-$ sector exhibits a similar pattern, with poles identified at approximately 2850--2910 MeV and 3100--3190 MeV. Interestingly, there is a physical state $D_{s1}(2860)$ corresponding to the lower pole in the $J^P = 1^-$ sector, which may be interpreted as a $D_1 K$ hadronic molecule.

Across all sectors, the implementation of Gaussian regularization tends to increase the widths of the higher poles, indicating sensitivity to short-distance dynamics. The predicted higher states generally have non-negligible widths and could potentially decay into final states such as $DK\pi$ and $DK\pi\pi$, making them candidates for future experimental studies.

Our results suggest a rich spectrum of charm-strange mesons that incorporates both molecular and quark model components, though further investigations including additional coupled channels and refined treatments of short-distance physics would be valuable to solidify these findings.

\begin{acknowledgments}
We would like to thank Guang-Juan Wang, Feng-Kun Guo and Xian-Hui Zhong for helpful discussions. This work is supported by the National Natural Science Foundation of China under Grants  No.~12235018, No.~11975165, No. 12305096, and the Fundamental Research Funds for the Central Universities under Grant No. SWU-KQ25016.
\end{acknowledgments}

\begin{appendix}
\begin{widetext}

\section{Chiral quark model}
\label{cqm}

In the chiral quark model~\cite{Manohar:1983md}, the low-energy interactions between quarks and NGBs in the $\mathrm{SU}(3)$ flavor basis are described by the effective Lagrangian~\cite{Li:1997gd, Zhao:2002id}
\begin{equation}
    \mathcal{L}_{Pqq}=\sum_j\frac{1}{f_m}\bar{\psi}_j\gamma_{\mu}^j\gamma_5^j\psi_j\partial^{\mu}\phi_m,
\end{equation}
where $\psi_j$ represents the $j$-th quark field in the hadron, $\phi_m$ denotes the NGB field, and $f_m$ is the corresponding decay constant.

We employ a harmonic-oscillator approximation for the spatial wave functions, where the linear confinement potential is neglected and the one-gluon exchange potential, $-\frac{4}{3}\frac{\alpha_s}{r}$, serves as the effective oscillator potential in the low-energy regime. Solving the Schrödinger equation yields spatial wave functions of the form
\begin{equation}
    \Psi_{nlm}(\bm{r})=R_{nl}(r)Y_{lm}(\theta,\phi),
\end{equation}
with the radial component given by
\begin{equation}
    R_{nl}(r)=\alpha^{\frac{3}{2}}\sqrt{\frac{2n!}{\Gamma(n+l+\frac{3}{2})}}(\alpha r)^l e^{-\alpha^2 r^2/2}L_n^{l+\frac{1}{2}}(\alpha^2 r^2),
\end{equation}
where $L_n^{l+\frac{1}{2}}$ represents an associated Laguerre polynomial. The oscillator parameter $\alpha$ incorporates the Hamiltonian parameters through the relation
\begin{equation}
    \alpha^2=\sqrt{\frac{2m_2}{m_1+m_2}}\beta^2.
\end{equation}

To maintain consistency with the nonrelativistic wave functions of heavy-light mesons, we utilize the nonrelativistic reduction of the Lagrangian. The Hamiltonian describing quark-NGB interactions within this framework reads~\cite{Li:1997gd, Zhao:2002id}
\begin{equation}
    \begin{aligned}
        H_m=\sum\limits_{j}\left\{\frac{\omega_m}{E_f+M_f}\bm{\sigma_j}\cdot \bm{P_f}+\frac{\omega_m}{E_i+M_i}\bm{\sigma_j}\cdot \bm{P_i}
        -\bm{\sigma_j}\cdot\bm{q}+\frac{\omega_m}{2\mu_q}\bm{\sigma_j}\cdot\bm{p'_j}\right\}I_j\varphi_m,
    \end{aligned}
\end{equation}
where $\bm{\sigma}_j$ denotes the Pauli spin operator acting on the $j$-th quark. The reduced mass $\mu_q$ is defined by $\mu_q^{-1} = m_j^{-1} + m_j'^{-1}$, with $m_j$ and $m_j'$ representing the masses of the $j$-th quark in the initial and final hadrons, respectively. The $I_j$ represents the flavor operator. The phase factor takes the form $\varphi_m = e^{-i\bm{q}\cdot\bm{r}_j}$ for meson emission and $\varphi_m = e^{i\bm{q}\cdot\bm{r}_j}$ for absorption. Here, $\bm{p}_j' = \bm{p}_j - \tfrac{m_j}{M}\bm{P}_{\text{c.m.}}$ corresponds to the internal momentum of the $j$-th quark in the rest frame of the initial meson, while $\omega_m$ and $\bm{q}$ denote the energy and three-momentum of the emitted NGB.

For strong decays involving light pseudoscalar meson emission from heavy-light mesons, the partial width is computed via
\begin{equation}
    \Gamma_{\chi\text{QM}}
    = \left(\frac{\delta}{f_m}\right)^2
      \frac{(E_f+M_f)|\bm{q}|}{4\pi M_i(2J_i+1)}
      \sum_{J_{iz},\,J_{fz}}
      \left|\mathcal{M}_{J_{iz},J_{fz}}\right|^2,
\end{equation}
where $\mathcal{M}$ represents the transition amplitude, and $J_{iz}$, $J_{fz}$ indicate the third components of the total angular momenta for the initial and final heavy-light mesons, respectively. The parameter $\delta$ serves as a universal coupling strength for quark-meson interactions.

In our numerical calculations, we adopt the following parameter values: $\beta = 0.4~\text{GeV}$, $\delta = 0.557$, and constituent quark masses $m_{u,d} = 0.35~\text{GeV}$, $m_s = 0.55~\text{GeV}$, $m_c = 1.7~\text{GeV}$. For the decay constants, we employ $f_{\pi} = 132~\text{MeV}$ and $f_{K,\eta} = 160~\text{MeV}$, consistent with Ref.~\cite{Zhong:2007gp}.

\end{widetext}
\end{appendix}


\begin{thebibliography}{99}

\bibitem{E598:1974sol}
J.~J.~Aubert \textit{et al.} [E598],
Phys. Rev. Lett. \textbf{33}, 1404-1406 (1974)
doi:10.1103/PhysRevLett.33.1404

\bibitem{SLAC-SP-017:1974ind}
J.~E.~Augustin \textit{et al.} [SLAC-SP-017],
Phys. Rev. Lett. \textbf{33}, 1406-1408 (1974)
doi:10.1103/PhysRevLett.33.1406

\bibitem{Eichten:1974af}
E.~Eichten, K.~Gottfried, T.~Kinoshita, J.~B.~Kogut, K.~D.~Lane and T.~M.~Yan,
Phys. Rev. Lett. \textbf{34}, 369-372 (1975)
[erratum: Phys. Rev. Lett. \textbf{36}, 1276 (1976)]
doi:10.1103/PhysRevLett.34.369

\bibitem{Godfrey:1985xj}
S.~Godfrey and N.~Isgur,
Phys. Rev. D \textbf{32}, 189-231 (1985)
doi:10.1103/PhysRevD.32.189

\bibitem{Capstick:1986ter}
S.~Capstick and N.~Isgur,
Phys. Rev. D \textbf{34}, no.9, 2809-2835 (1986)
doi:10.1103/physrevd.34.2809

\bibitem{Godfrey:1986wj}
S.~Godfrey and R.~Kokoski,
Phys. Rev. D \textbf{43}, 1679-1687 (1991)
doi:10.1103/PhysRevD.43.1679

\bibitem{BaBar:2003oey}
B.~Aubert \textit{et al.} [BaBar],
Phys. Rev. Lett. \textbf{90}, 242001 (2003)
doi:10.1103/PhysRevLett.90.242001
[arXiv:hep-ex/0304021 [hep-ex]].

\bibitem{CLEO:2003ggt}
D.~Besson \textit{et al.} [CLEO],
Phys. Rev. D \textbf{68}, 032002 (2003)
[erratum: Phys. Rev. D \textbf{75}, 119908 (2007)]
doi:10.1103/PhysRevD.68.032002
[arXiv:hep-ex/0305100 [hep-ex]].

\bibitem{Guo:2017jvc}
F.~K.~Guo, C.~Hanhart, U.~G.~Mei{\ss}ner, Q.~Wang, Q.~Zhao and B.~S.~Zou,
Rev. Mod. Phys. \textbf{90}, no.1, 015004 (2018)
[erratum: Rev. Mod. Phys. \textbf{94}, no.2, 029901 (2022)]
doi:10.1103/RevModPhys.90.015004
[arXiv:1705.00141 [hep-ph]].

\bibitem{Olsen:2017bmm}
S.~L.~Olsen, T.~Skwarnicki and D.~Zieminska,
Rev. Mod. Phys. \textbf{90}, no.1, 015003 (2018)
doi:10.1103/RevModPhys.90.015003
[arXiv:1708.04012 [hep-ph]].

\bibitem{Liu:2019zoy}
Y.~R.~Liu, H.~X.~Chen, W.~Chen, X.~Liu and S.~L.~Zhu,
Prog. Part. Nucl. Phys. \textbf{107}, 237-320 (2019)
doi:10.1016/j.ppnp.2019.04.003
[arXiv:1903.11976 [hep-ph]].

\bibitem{Chen:2016qju}
H.~X.~Chen, W.~Chen, X.~Liu and S.~L.~Zhu,
Phys. Rept. \textbf{639}, 1-121 (2016)
doi:10.1016/j.physrep.2016.05.004
[arXiv:1601.02092 [hep-ph]].

\bibitem{Liu:2012zya}
L.~Liu, K.~Orginos, F.~K.~Guo, C.~Hanhart and U.~G.~Meissner,
Phys. Rev. D \textbf{87}, no.1, 014508 (2013)
doi:10.1103/PhysRevD.87.014508
[arXiv:1208.4535 [hep-lat]].

\bibitem{Mohler:2013rwa}
D.~Mohler, C.~B.~Lang, L.~Leskovec, S.~Prelovsek and R.~M.~Woloshyn,
Phys. Rev. Lett. \textbf{111}, no.22, 222001 (2013)
doi:10.1103/PhysRevLett.111.222001
[arXiv:1308.3175 [hep-lat]].

\bibitem{Belle-II:2025dzk}
M.~Abumusabh \textit{et al.} [Belle-II],
[arXiv:2510.27174 [hep-ex]].

\bibitem{Fu:2021wde}
H.~L.~Fu, H.~W.~Grie{\ss}hammer, F.~K.~Guo, C.~Hanhart and U.~G.~Mei{\ss}ner,
Eur. Phys. J. A \textbf{58}, no.4, 70 (2022)
doi:10.1140/epja/s10050-022-00724-8
[arXiv:2111.09481 [hep-ph]].

\bibitem{Bicudo:2005de}
P.~Bicudo,
Phys. Rev. D \textbf{74}, 036008 (2006)
doi:10.1103/PhysRevD.74.036008
[arXiv:hep-ph/0512041 [hep-ph]].

\bibitem{MartinezTorres:2011pr}
A.~Martinez Torres, L.~R.~Dai, C.~Koren, D.~Jido and E.~Oset,
Phys. Rev. D \textbf{85}, 014027 (2012)
doi:10.1103/PhysRevD.85.014027
[arXiv:1109.0396 [hep-lat]].

\bibitem{MartinezTorres:2014kpc}
A.~Mart{\'\i}nez Torres, E.~Oset, S.~Prelovsek and A.~Ramos,
JHEP \textbf{05}, 153 (2015)
doi:10.1007/JHEP05(2015)153
[arXiv:1412.1706 [hep-lat]].

\bibitem{Ortega:2016mms}
P.~G.~Ortega, J.~Segovia, D.~R.~Entem and F.~Fernandez,
Phys. Rev. D \textbf{94}, no.7, 074037 (2016)
doi:10.1103/PhysRevD.94.074037
[arXiv:1603.07000 [hep-ph]].

\bibitem{Albaladejo:2018mhb}
M.~Albaladejo, P.~Fernandez-Soler, J.~Nieves and P.~G.~Ortega,
Eur. Phys. J. C \textbf{78}, no.9, 722 (2018)
doi:10.1140/epjc/s10052-018-6176-3
[arXiv:1805.07104 [hep-ph]].

\bibitem{Yang:2021tvc}
Z.~Yang, G.~J.~Wang, J.~J.~Wu, M.~Oka and S.~L.~Zhu,
Phys. Rev. Lett. \textbf{128}, no.11, 112001 (2022)
doi:10.1103/PhysRevLett.128.112001
[arXiv:2107.04860 [hep-ph]].

\bibitem{Zhang:2024usz}
Z.~L.~Zhang, Z.~W.~Liu, S.~Q.~Luo, P.~Chen and Z.~H.~Guo,
Phys. Rev. D \textbf{110}, no.9, 094037 (2024)
doi:10.1103/PhysRevD.110.094037
[arXiv:2409.05337 [hep-ph]].

\bibitem{Shen:2025qpj}
Y.~b.~Shen, Z.~W.~Liu, J.~X.~Lu, M.~Z.~Liu and L.~S.~Geng,
[arXiv:2506.23476 [hep-ph]].

\bibitem{Zhou:2020moj}
Z.~Y.~Zhou and Z.~Xiao,
Eur. Phys. J. C \textbf{81}, no.6, 551 (2021)
doi:10.1140/epjc/s10052-021-09329-9
[arXiv:2008.08002 [hep-ph]].

\bibitem{Zeng:1994vj}
J.~Zeng, J.~W.~Van Orden and W.~Roberts,
Phys. Rev. D \textbf{52}, 5229-5241 (1995)
doi:10.1103/PhysRevD.52.5229
[arXiv:hep-ph/9412269 [hep-ph]].

\bibitem{Lahde:1999ih}
T.~A.~Lahde, C.~J.~Nyfalt and D.~O.~Riska,
Nucl. Phys. A \textbf{674}, 141-167 (2000)
doi:10.1016/S0375-9474(00)00154-8
[arXiv:hep-ph/9908485 [hep-ph]].

\bibitem{DiPierro:2001dwf}
M.~Di Pierro and E.~Eichten,
Phys. Rev. D \textbf{64}, 114004 (2001)
doi:10.1103/PhysRevD.64.114004
[arXiv:hep-ph/0104208 [hep-ph]].

\bibitem{Ebert:2009ua}
D.~Ebert, R.~N.~Faustov and V.~O.~Galkin,
Eur. Phys. J. C \textbf{66}, 197-206 (2010)
doi:10.1140/epjc/s10052-010-1233-6
[arXiv:0910.5612 [hep-ph]].

\bibitem{Li:2010vx}
D.~M.~Li, P.~F.~Ji and B.~Ma,
Eur. Phys. J. C \textbf{71}, 1582 (2011)
doi:10.1140/epjc/s10052-011-1582-9
[arXiv:1011.1548 [hep-ph]].

\bibitem{Godfrey:2015dva}
S.~Godfrey and K.~Moats,
Phys. Rev. D \textbf{93}, no.3, 034035 (2016)
doi:10.1103/PhysRevD.93.034035
[arXiv:1510.08305 [hep-ph]].

\bibitem{Ni:2023lvx}
R.~H.~Ni, J.~J.~Wu and X.~H.~Zhong,
Phys. Rev. D \textbf{109}, no.11, 116006 (2024)
doi:10.1103/PhysRevD.109.116006
[arXiv:2312.04765 [hep-ph]].

\bibitem{ParticleDataGroup:2024cfk}
S.~Navas \textit{et al.} [Particle Data Group],
Phys. Rev. D \textbf{110}, no.3, 030001 (2024)
doi:10.1103/PhysRevD.110.030001

\bibitem{Wise:1992hn}
M.~B.~Wise,
Phys. Rev. D \textbf{45}, no.7, R2188 (1992)
doi:10.1103/PhysRevD.45.R2188

\bibitem{Casalbuoni:1996pg}
R.~Casalbuoni, A.~Deandrea, N.~Di Bartolomeo, R.~Gatto, F.~Feruglio and G.~Nardulli,
Phys. Rept. \textbf{281}, 145-238 (1997)
doi:10.1016/S0370-1573(96)00027-0
[arXiv:hep-ph/9605342 [hep-ph]].

\bibitem{Abreu:2011ic}
L.~M.~Abreu, D.~Cabrera, F.~J.~Llanes-Estrada and J.~M.~Torres-Rincon,
Annals Phys. \textbf{326}, 2737-2772 (2011)
doi:10.1016/j.aop.2011.06.006
[arXiv:1104.3815 [hep-ph]].

\bibitem{Hyodo:2006kg}
T.~Hyodo, D.~Jido and A.~Hosaka,
Phys. Rev. D \textbf{75}, 034002 (2007)
doi:10.1103/PhysRevD.75.034002
[arXiv:hep-ph/0611004 [hep-ph]].

\bibitem{Cahn:2003cw}
R.~N.~Cahn and J.~D.~Jackson,
Phys. Rev. D \textbf{68}, 037502 (2003)
doi:10.1103/PhysRevD.68.037502
[arXiv:hep-ph/0305012 [hep-ph]].

\bibitem{Pelaez:2015qba}
J.~R.~Pelaez,
Phys. Rept. \textbf{658}, 1 (2016)
doi:10.1016/j.physrep.2016.09.001
[arXiv:1510.00653 [hep-ph]].

\bibitem{Oset:1997it}
E.~Oset and A.~Ramos,
Nucl. Phys. A \textbf{635}, 99-120 (1998)
doi:10.1016/S0375-9474(98)00170-5
[arXiv:nucl-th/9711022 [nucl-th]].

\bibitem{Oller:1998hw}
J.~A.~Oller, E.~Oset and J.~R.~Pelaez,
Phys. Rev. D \textbf{59}, 074001 (1999)
[erratum: Phys. Rev. D \textbf{60}, 099906 (1999); erratum: Phys. Rev. D \textbf{75}, 099903 (2007)]
doi:10.1103/PhysRevD.59.074001
[arXiv:hep-ph/9804209 [hep-ph]].

\bibitem{Weinberg:1962hj}
S.~Weinberg,
Phys. Rev. \textbf{130}, 776-783 (1963)
doi:10.1103/PhysRev.130.776

\bibitem{Baru:2003qq}
V.~Baru, J.~Haidenbauer, C.~Hanhart, Y.~Kalashnikova and A.~E.~Kudryavtsev,
Phys. Lett. B \textbf{586}, 53-61 (2004)
doi:10.1016/j.physletb.2004.01.088
[arXiv:hep-ph/0308129 [hep-ph]].

\bibitem{Gamermann:2009uq}
D.~Gamermann, J.~Nieves, E.~Oset and E.~Ruiz Arriola,
Phys. Rev. D \textbf{81}, 014029 (2010)
doi:10.1103/PhysRevD.81.014029
[arXiv:0911.4407 [hep-ph]].

\bibitem{Aceti:2014ala}
F.~Aceti, L.~R.~Dai, L.~S.~Geng, E.~Oset and Y.~Zhang,
Eur. Phys. J. A \textbf{50}, 57 (2014)
doi:10.1140/epja/i2014-14057-2
[arXiv:1301.2554 [hep-ph]].

\bibitem{Aceti:2012dd}
F.~Aceti and E.~Oset,
Phys. Rev. D \textbf{86}, 014012 (2012)
doi:10.1103/PhysRevD.86.014012
[arXiv:1202.4607 [hep-ph]].

\bibitem{Hofmann:2003je}
J.~Hofmann and M.~F.~M.~Lutz,
Nucl. Phys. A \textbf{733}, 142-152 (2004)
doi:10.1016/j.nuclphysa.2003.12.013
[arXiv:hep-ph/0308263 [hep-ph]].

\bibitem{Guo:2006fu}
F.~K.~Guo, P.~N.~Shen, H.~C.~Chiang, R.~G.~Ping and B.~S.~Zou,
Phys. Lett. B \textbf{641}, 278-285 (2006)
doi:10.1016/j.physletb.2006.08.064
[arXiv:hep-ph/0603072 [hep-ph]].

\bibitem{Wang:2006zw}
Z.~G.~Wang,
J. Phys. G \textbf{34}, 753-765 (2007)
doi:10.1088/0954-3899/34/4/011
[arXiv:hep-ph/0611271 [hep-ph]].

\bibitem{Faessler:2007gv}
A.~Faessler, T.~Gutsche, V.~E.~Lyubovitskij and Y.~L.~Ma,
Phys. Rev. D \textbf{76}, 014005 (2007)
doi:10.1103/PhysRevD.76.014005
[arXiv:0705.0254 [hep-ph]].

\bibitem{Guo:2008gp}
F.~K.~Guo, C.~Hanhart, S.~Krewald and U.~G.~Meissner,
Phys. Lett. B \textbf{666}, 251-255 (2008)
doi:10.1016/j.physletb.2008.07.060
[arXiv:0806.3374 [hep-ph]].

\bibitem{Guo:2009ct}
F.~K.~Guo, C.~Hanhart and U.~G.~Meissner,
Eur. Phys. J. A \textbf{40}, 171-179 (2009)
doi:10.1140/epja/i2009-10762-1
[arXiv:0901.1597 [hep-ph]].

\bibitem{Cleven:2014oka}
M.~Cleven, H.~W.~Grie{\ss}hammer, F.~K.~Guo, C.~Hanhart and U.~G.~Mei{\ss}ner,
Eur. Phys. J. A \textbf{50}, 149 (2014)
doi:10.1140/epja/i2014-14149-y
[arXiv:1405.2242 [hep-ph]].

\bibitem{Guo:2015dha}
Z.~H.~Guo, U.~G.~Mei{\ss}ner and D.~L.~Yao,
Phys. Rev. D \textbf{92}, no.9, 094008 (2015)
doi:10.1103/PhysRevD.92.094008
[arXiv:1507.03123 [hep-ph]].

\bibitem{Albaladejo:2016hae}
M.~Albaladejo, D.~Jido, J.~Nieves and E.~Oset,
Eur. Phys. J. C \textbf{76}, no.6, 300 (2016)
doi:10.1140/epjc/s10052-016-4144-3
[arXiv:1604.01193 [hep-ph]].

\bibitem{Du:2017ttu}
M.~L.~Du, F.~K.~Guo, U.~G.~Mei{\ss}ner and D.~L.~Yao,
Eur. Phys. J. C \textbf{77}, no.11, 728 (2017)
doi:10.1140/epjc/s10052-017-5287-6
[arXiv:1703.10836 [hep-ph]].

\bibitem{Gil-Dominguez:2023puj}
F.~Gil-Dom{\'\i}nguez and R.~Molina,
Phys. Rev. D \textbf{109}, no.9, 096002 (2024)
doi:10.1103/PhysRevD.109.096002
[arXiv:2306.01848 [hep-ph]].

\bibitem{Yan:2018zdt}
M.~J.~Yan, X.~H.~Liu, S.~Gonz{\`a}lez-Sol{\'\i}s, F.~K.~Guo, C.~Hanhart, U.~G.~Mei{\ss}ner and B.~S.~Zou,
Phys. Rev. D \textbf{98}, no.9, 091502 (2018)
doi:10.1103/PhysRevD.98.091502
[arXiv:1805.10972 [hep-ph]].

\bibitem{Peng:2020gwk}
F.~Z.~Peng, J.~X.~Lu, M.~S{\'a}nchez S{\'a}nchez, M.~J.~Yan and M.~Pavon Valderrama,
Phys. Rev. D \textbf{103}, no.1, 014023 (2021)
doi:10.1103/PhysRevD.103.014023
[arXiv:2007.01198 [hep-ph]].

\bibitem{Wu:2024bvl}
J.~Z.~Wu, J.~Y.~Pang and J.~J.~Wu,
Chin. Phys. Lett. \textbf{41}, no.9, 091201 (2024)
doi:10.1088/0256-307X/41/9/091201
[arXiv:2407.05743 [hep-ph]].

\bibitem{LHCb:2014ioa}
R.~Aaij \textit{et al.} [LHCb],
Phys. Rev. D \textbf{90}, no.7, 072003 (2014)
doi:10.1103/PhysRevD.90.072003
[arXiv:1407.7712 [hep-ex]].

\bibitem{BaBar:2014jjr}
J.~P.~Lees \textit{et al.} [BaBar],
Phys. Rev. D \textbf{91}, no.5, 052002 (2015)
doi:10.1103/PhysRevD.91.052002
[arXiv:1412.6751 [hep-ex]].

\bibitem{LHCb:2018oeb}
R.~Aaij \textit{et al.} [LHCb],
Phys. Rev. D \textbf{98}, no.7, 072006 (2018)
doi:10.1103/PhysRevD.98.072006
[arXiv:1807.01891 [hep-ex]].

\bibitem{Guo:2006rp}
F.~K.~Guo, P.~N.~Shen and H.~C.~Chiang,
Phys. Lett. B \textbf{647}, 133-139 (2007)
doi:10.1016/j.physletb.2007.01.050
[arXiv:hep-ph/0610008 [hep-ph]].

\bibitem{Gamermann:2007fi}
D.~Gamermann and E.~Oset,
Eur. Phys. J. A \textbf{33}, 119-131 (2007)
doi:10.1140/epja/i2007-10435-1
[arXiv:0704.2314 [hep-ph]].

\bibitem{Faessler:2007us}
A.~Faessler, T.~Gutsche, V.~E.~Lyubovitskij and Y.~L.~Ma,
Phys. Rev. D \textbf{76}, 114008 (2007)
doi:10.1103/PhysRevD.76.114008
[arXiv:0709.3946 [hep-ph]].

\bibitem{Cleven:2010aw}
M.~Cleven, F.~K.~Guo, C.~Hanhart and U.~G.~Meissner,
Eur. Phys. J. A \textbf{47}, 19 (2011)
doi:10.1140/epja/i2011-11019-2
[arXiv:1009.3804 [hep-ph]].

\bibitem{Du:2017zvv}
M.~L.~Du, M.~Albaladejo, P.~Fern{\'a}ndez-Soler, F.~K.~Guo, C.~Hanhart, U.~G.~Mei{\ss}ner, J.~Nieves and D.~L.~Yao,
Phys. Rev. D \textbf{98}, no.9, 094018 (2018)
doi:10.1103/PhysRevD.98.094018
[arXiv:1712.07957 [hep-ph]].

\bibitem{BaBar:2006gme}
B.~Aubert \textit{et al.} [BaBar],
Phys. Rev. Lett. \textbf{97}, 222001 (2006)
doi:10.1103/PhysRevLett.97.222001
[arXiv:hep-ex/0607082 [hep-ex]].

\bibitem{BaBar:2009rro}
B.~Aubert \textit{et al.} [BaBar],
Phys. Rev. D \textbf{80}, 092003 (2009)
doi:10.1103/PhysRevD.80.092003
[arXiv:0908.0806 [hep-ex]].

\bibitem{LHCb:2012uts}
R.~Aaij \textit{et al.} [LHCb],
JHEP \textbf{10}, 151 (2012)
doi:10.1007/JHEP10(2012)151
[arXiv:1207.6016 [hep-ex]].

\bibitem{LHCb:2014ott}
R.~Aaij \textit{et al.} [LHCb],
Phys. Rev. Lett. \textbf{113}, 162001 (2014)
doi:10.1103/PhysRevLett.113.162001
[arXiv:1407.7574 [hep-ex]].

\bibitem{Chen:2016spr}
H.~X.~Chen, W.~Chen, X.~Liu, Y.~R.~Liu and S.~L.~Zhu,
Rept. Prog. Phys. \textbf{80}, no.7, 076201 (2017)
doi:10.1088/1361-6633/aa6420
[arXiv:1609.08928 [hep-ph]].

\bibitem{Manohar:1983md}
A.~Manohar and H.~Georgi,
Nucl. Phys. B \textbf{234}, 189-212 (1984)
doi:10.1016/0550-3213(84)90231-1

\bibitem{Li:1997gd}
Z.~p.~Li, H.~x.~Ye and M.~h.~Lu,
Phys. Rev. C \textbf{56}, 1099-1113 (1997)
doi:10.1103/PhysRevC.56.1099
[arXiv:nucl-th/9706010 [nucl-th]].

\bibitem{Zhao:2002id}
Q.~Zhao, J.~S.~Al-Khalili, Z.~P.~Li and R.~L.~Workman,
Phys. Rev. C \textbf{65}, 065204 (2002)
doi:10.1103/PhysRevC.65.065204
[arXiv:nucl-th/0202067 [nucl-th]].

\bibitem{Zhong:2007gp}
X.~H.~Zhong and Q.~Zhao,
Phys. Rev. D \textbf{77}, 074008 (2008)
doi:10.1103/PhysRevD.77.074008
[arXiv:0711.4645 [hep-ph]].

\end{thebibliography}

\end{document}